\documentclass[review]{elsarticle}

\usepackage{lineno,hyperref}
\modulolinenumbers[5]

\newcommand{\dd}{{\rm d}}
\newcommand{\pp}{\tilde{p}}
\newcommand{\ee}{\tilde{\varepsilon}}

\usepackage{xfrac}
\DeclareMathAlphabet{\mathpzc}{OT1}{pzc}{m}{it}

\usepackage[dvipsnames]{xcolor}
\usepackage{mathrsfs}
\DeclareMathAlphabet{\mathrsfso}{T1}{cm}{m}{n} 
\DeclareMathAlphabet{\mathcalligra}{T1}{calligra}{m}{n}

\journal{Journal of \LaTeX\ Templates}

\begin{document}

\begin{frontmatter}

\title{Spectral features of non-equilibrium antineutrinos of primordial nucleosynthesis}


\author[mymainaddress]{Vlad Yu. Yurchenko}
\ead{yurchvlad@gmail.com}

\author[mymainaddress]{Alexandre V. Ivanchik\corref{mycorrespondingauthor}}
\cortext[mycorrespondingauthor]{Corresponding author}
\ead{iav@astro.ioffe.ru}

\address[mymainaddress]{Ioffe Institute, Politekhnicheskaya 26, 194021 St.-Petersburg, Russia}

\begin{abstract}
During the era of primordial nucleosynthesis the background of non-equilibrium antineutrinos is being formed due to decays of neutrons and nuclei of tritium.
The spectra of antineutrinos of this background were calculated taking into account the Coulomb interaction between electron and daughter nucleus in $\beta^\text{--}$-decay.
The dependence of these spectra on the value of the baryon-to-photon ratio $\eta$ at the period of primordial nucleosynthesis is investigated.
The observations of these antineutrinos will allow us to look directly at the very early Universe and nonequilibrium processes taken place before, during, and some time after primordial nucleosynthesis.
In any case, this phenomenon is one more aspect in the picture of the standard cosmological model.
\end{abstract}


\end{frontmatter}


\flushbottom

\section{Introduction}
\label{sec:introduction}

The observations of the cosmic microwave background (CMB) allow us to see into our Universe when it was about 380 000 years old.
Similarly primordial nucleosynthesis provides us with an indirect probe of the early Universe (about a few minutes old) based on the comparison of light element (D, $^4$He, $^7$Li) observations with corresponding theoretical calculations, which in turn is based on the well-established knowledge of nuclear and particle physics \cite{W2008, GR2018, KT1990, P1993}.
We cannot observe the Universe at that epoch directly using electromagnetic radiation due to the opacity of the Universe at early stages right up until primordial recombination. 
Nevertheless, the direct information about the first seconds of the Universe evolution principally can be obtained by the detection of relic neutrinos which were going out of equilibrium and beginning a free expansion without any interaction with the other primordial material when the age of the Universe was less then one second.
These neutrinos as well known as the cosmic neutrino background C$\nu$B (note that cosmic neutrino background consists of neutrino as well as antineutrino, so it is more correct to use the following abbreviation C$\nu\tilde{\nu}$B but we use commonly accepted C$\nu$B).
Like the cosmic microwave background radiation, the C$\nu$B was formed with a thermal equilibrium spectrum which for neutrinos ($\nu$) and antineutrinos ($\tilde{\nu}$) is given by the Fermi-Dirac distribution:
\begin{equation}
\label{eq:eq_mom_du_fu}
n_{\nu\tilde{\nu}}(p)dp=\frac{1}{(2\pi\hbar)^3}\frac{4\pi p^2dp}{\exp(pc/kT)+1}.
\end{equation}

This formula is expressed for the massless antineutrino case ($\varepsilon_{\nu}=pc$).
We can neglect the neutrino mass at the decoupling period (T $\sim$ 2 MeV) because there is the upper limit of $\sum m_{\nu}<0.23$ eV \cite{TAL2010, PC2016} and the ratio $m_{\nu}c^2/kT$ is about 10$^{-7}$ for this period.
After neutrino decoupling the spectrum has kept the same form (due to the adiabatic expansion of the Universe) with a temperature decreasing like $T_{\nu}\propto(1+z)$, where $z$ is the cosmological redshift.
It is very important to note that this fact takes place for momentum distribution $n(p)$ whether or not neutrino possesses mass, while the form of energy distribution $n(\varepsilon)$ depends on neutrino mass.
Therefore despite the fact that $m_{\nu}\neq0$, today ($z{=}0$) the momentum distribution $n(p)$ has the form to be the same as eq. \ref{eq:eq_mom_du_fu} with the current temperature $T_{\nu0}$ whose value is related to the current temperature of the relic photons $T_{\gamma0}$ (the CMB temperature).
The thermodynamics of the early Universe give us the relation between the relic neutrino and photon temperatures, $T_{\nu}=(4/11)^{1/3}T_{\gamma}$, arising from electron-positron annihilation.
Given this relation and the current value of $T_{\gamma0}\approx2.725\pm0.001$ K \cite{F2009}, we have the present temperature of relic neutrinos to be $T_{\nu}\approx1.945$ K.
However, small entropy transferring from electrons and positrons into neutrinos and antineutrinos during the epoch of $e^-e^+$--annihilation leads to the minor energy-dependent distortions of relic neutrinos energy spectrum \cite{DF1992,DT1992,HM1995,DHS1997}, therefore, strictly speaking, the whole spectrum cannot be described by only a single parameter $T_{\nu0}$.

The equilibrium part of relic neutrinos (C$\nu$B) carries only the information about thermal equilibrium between neutrinos and electron-nucleon-photon plasma before neutrino decoupling.
In contrast distortions of relic neutrinos carry information about non-equilibrium processes with the participation of neutrinos after decoupling.
Besides distortions due to $e^-e^+$--annihilation mentioned above there is one more significant non-equilibrium addition to C$\nu$B formed by antineutrinos which were being produced during the era of Big Bang nucleosynthesis (BBN) as a result of decays of $\beta^-$ unstable neutrons n and nuclei of tritium (tritons) t \cite{IY2018} (mass fractions of neutrons and tritons during primordial nucleosynthesis are shown in figure \ref{fig:relative_mass_fractions}).
The spectra of these antineutrinos are non-equilibrium because the period of their formation take place rather after neutrino decoupling.
These antineutrino spectra carry the information about the temporal evolution of abundances of these elements at that epoch.

\subsection{``Grand Unified Neutrino Spectrum''}

\begin{figure}[t]
	\centering 
	\includegraphics[width=1.\textwidth,origin=c,angle=0]{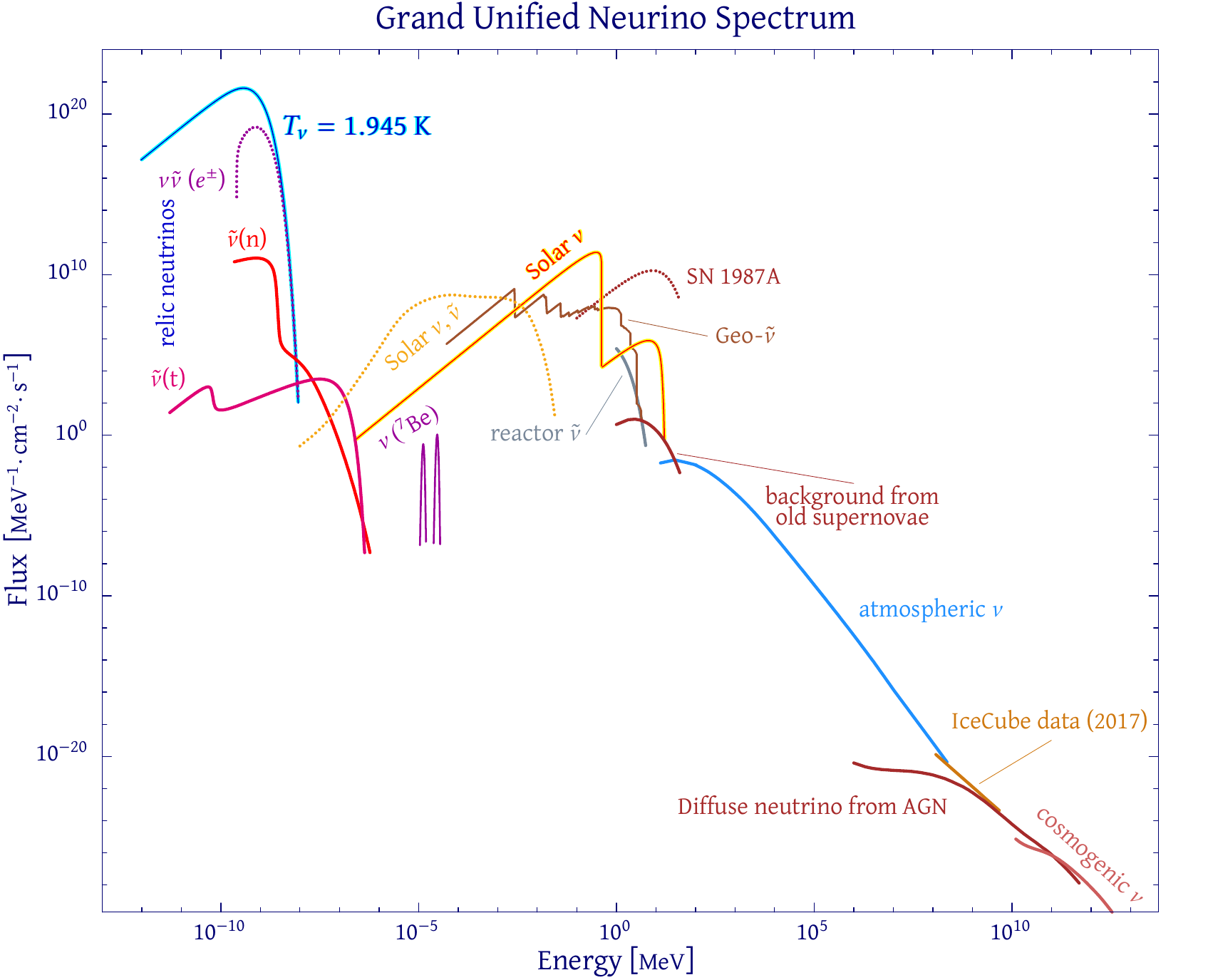}
	\caption{\label{fig:GUNS} The spectra of neutrinos and antineutrinos from different natural sources. Relic neutrinos and antineutrinos (C$\nu$B) with temperature $T_{\nu}=1.945$ K \cite{W2008, GR2018, KT1990, P1993}; a non-thermal addition to relic neutrinos and antineutrinos due to $e^+e^-$-annihilation at the beginning of primordial nucleosynthesis \cite{DF1992,DT1992,HM1995,DHS1997}; non-equilibrium antineutrinos due to neutrons and tritons decays during primordial nucleosynthesis \cite{IY2018}; solar neutrinos from reactions of thermonuclear synthesis \cite{BSB2005}; the calculated spectrum of solar neutrinos and antineutrinos at keV energies emerging from various thermal processes in the solar plasma \cite{VRR}; the doublet of the narrow cosmological neutrino lines from $^7$Be decay \cite{KS2011}; antineutrinos produced geologically (geoneutrinos) in decays of radioactive isotopes \cite{E2006, E2007, PhysRevD.99.093009}; the diffuse supernova neutrino background \cite{B2010}; the calculated spectrum of the neutrino burst from SN1987a detected on Earth surface \cite{H1987, B1987}; atmospheric neutrinos \cite{H2004}; theoretically  estimated neutrino flux from inner jets of AGN \cite{M2014}; IceCube data (2017) \cite{arXiv:1710}, sources of the most high energetic neutrinos detected by IceCube are still not fully known; cosmogenic neutrinos \cite{A2015}. For detailed information on high energy fluxes see \cite{VTR2019}.}
\end{figure}

In addition to relic neutrinos it should be mentioned that there are other neutrino backgrounds at higher energies which could be called ``cosmological'' as well: two narrow neutrino lines from ${^7}$Be decay \cite{KS2011} taken place after its recombination epoch, the diffuse supernova neutrino background \cite{B2010}, the active galactic nuclei (AGN) background and cosmogenic (GZK) neutrinos arising from interactions of ultraenergetic protons with the CMB photons (see e.g. \cite{KS2012} and references therein).

The spectra of C$\nu$B, energy-dependent admixture to these spectra due to $e^{\pm}$-annihilation and non-equilibrium spectra of ``neutron'' and ``triton'' antineutrino of primordial nucleosynthesis are shown in figure \ref{fig:GUNS}, which represent so called Grand Unified Neutrino Spectrum, among a number of neutrino/antineutrino spectra from other different natural sources.\footnote{The name ``Grand Unified Neutrino Spectrum'' was taken from \cite{VRR}.}
It can be seen that the spectra of relic neutrinos/antineutrinos occupy the area of the lowest energies in figure \ref{fig:GUNS}, because of the remoteness of the eras of their formation.

In this work we discuss the complementary ability to look at the early Universe and non-equilibrium processes occurred before, during and after primordial nucleosynthesis using spectrum calculations and future possible observations of antineutrinos having arisen from neutron and triton (nucleus of tritium) decays ($n\rightarrow p+e^-+\tilde{\nu}_e$, $t\rightarrow {^3}he+e^-+\tilde{\nu}_e$).

\section{Non-equilibrium antineutrino spectra}

\subsection{Free decaying nuclei at the BBN epoch}
\label{sec:decaying_nuclei}
The number of antineutrinos produced in decays of neutrons and tritons per unit volume during a time interval $\dd t$ with momentum between  $p$ and $p+\dd p$ is
\begin{equation}
\label{eq:number_of_particles}
\dd n_{\nu}(p)=\lambda n(t)f(p)\dd p\dd t,
\end{equation}
where $\lambda$ is the decay rate of radioactive nuclear species in question, $n(t)$ is its number density, $f(p)$ is the momentum spectrum of antineutrino produced in $\beta^-$ decay, it is normalized so that
\begin{equation}
\label{eq:momentum_distribution_normalizing}
\int\limits_0^{p_{max}}\!\!\!f(p)\dd p=1,
\end{equation}
where $p_{max}$ is the maximum antineutrino momentum in $\beta^-$ decay.
The number density $n(t)$ of neutrons or tritons can be expressed as $n(z)=Y(z)n_b(z)=Y(z)n^0_b(1+z)^3$, where $Y(z)$ is the ratio of neutrons or tritons to all baryons $n_b$, and $n^0_b=\eta n^0_{\gamma}\simeq2.48\times10^{-7}$ cm$^{-3}$ (see  e.g.  \cite{PDG2014}) is the baryon number density at the present epoch ($z=0$), $\eta $ is baryon-to-photon ratio at the present epoch and its value is $\eta=6.1\times10^{-10}$ \cite{PC2018}.
To obtain the dependencies $Y(t)$ we have updated our own previous numerical code for primordial nucleisynthesis \cite{OIV2000} which is based on the historical Wagoner's code \cite{WFH1967} (results of our code are in good agreement with presented results of other known codes (e.g. \cite{PCEIMMS2008, CPUVDIL2015, CFOY2016, PCUV2018}).

\begin{figure}[tbp]
	\centering 
	\includegraphics[width=1.\textwidth,trim=0 0 0 0, clip]{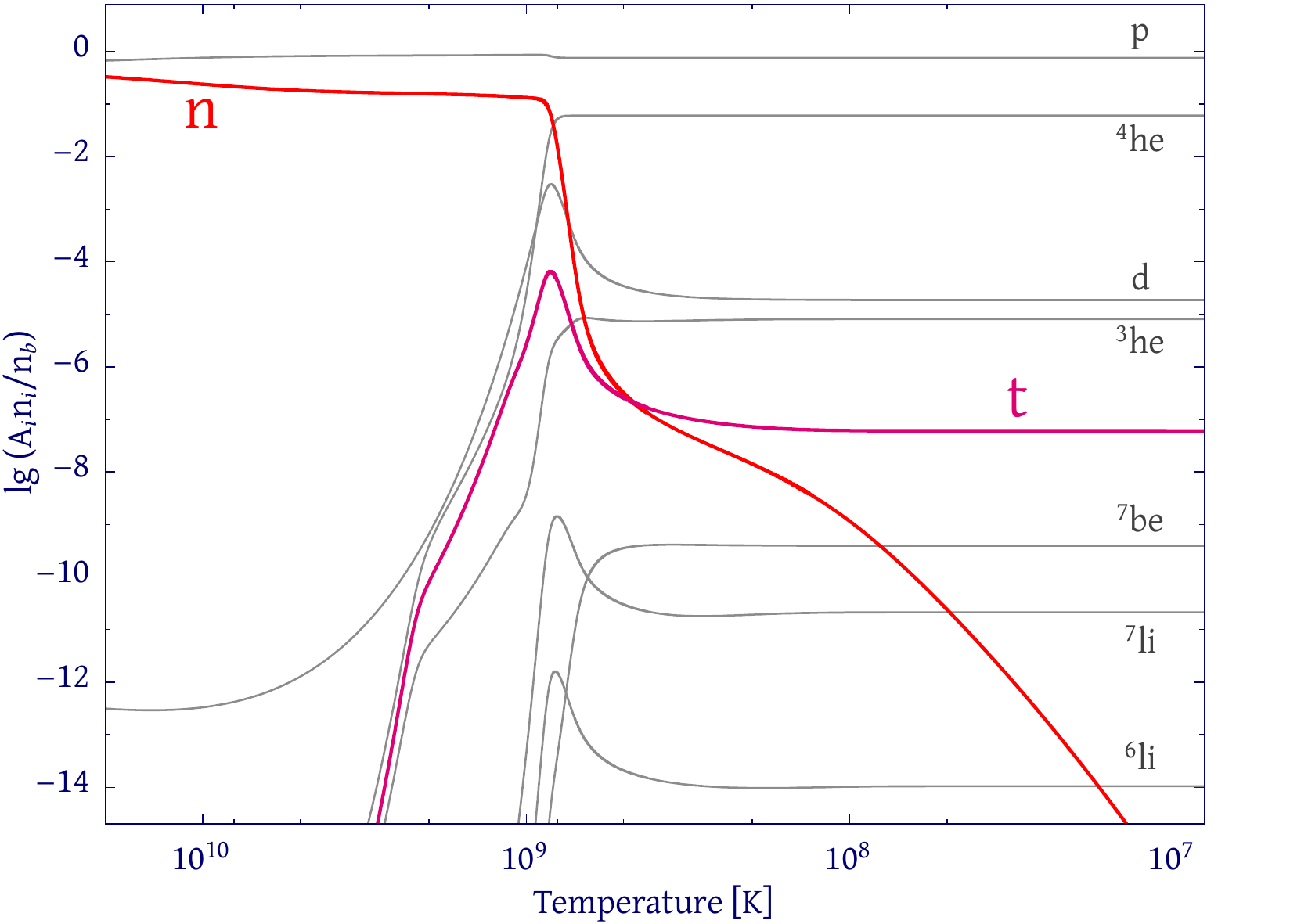}

	\caption{The evolution of the mass fractions $A_in_i/n_b$ ($A_i$ is the mass number of species $i$) of the light nuclei produced during primordial nucleosynthesis as a function of the temperature. Red and crimson lines correspond to neutron and tritium mass fractions. The calculations have been done by using our own numerical code for primordial nucleosynthesis (see the text) \cite{OIV2000}.}
	\label{fig:relative_mass_fractions}
	\hfill
\end{figure}

\subsection{Non-equilibrium antineutrino spectral characteristics}

\begin{figure}[tbp]
\centering 
\includegraphics[width=.95\textwidth,trim=0 0 0 0,clip]{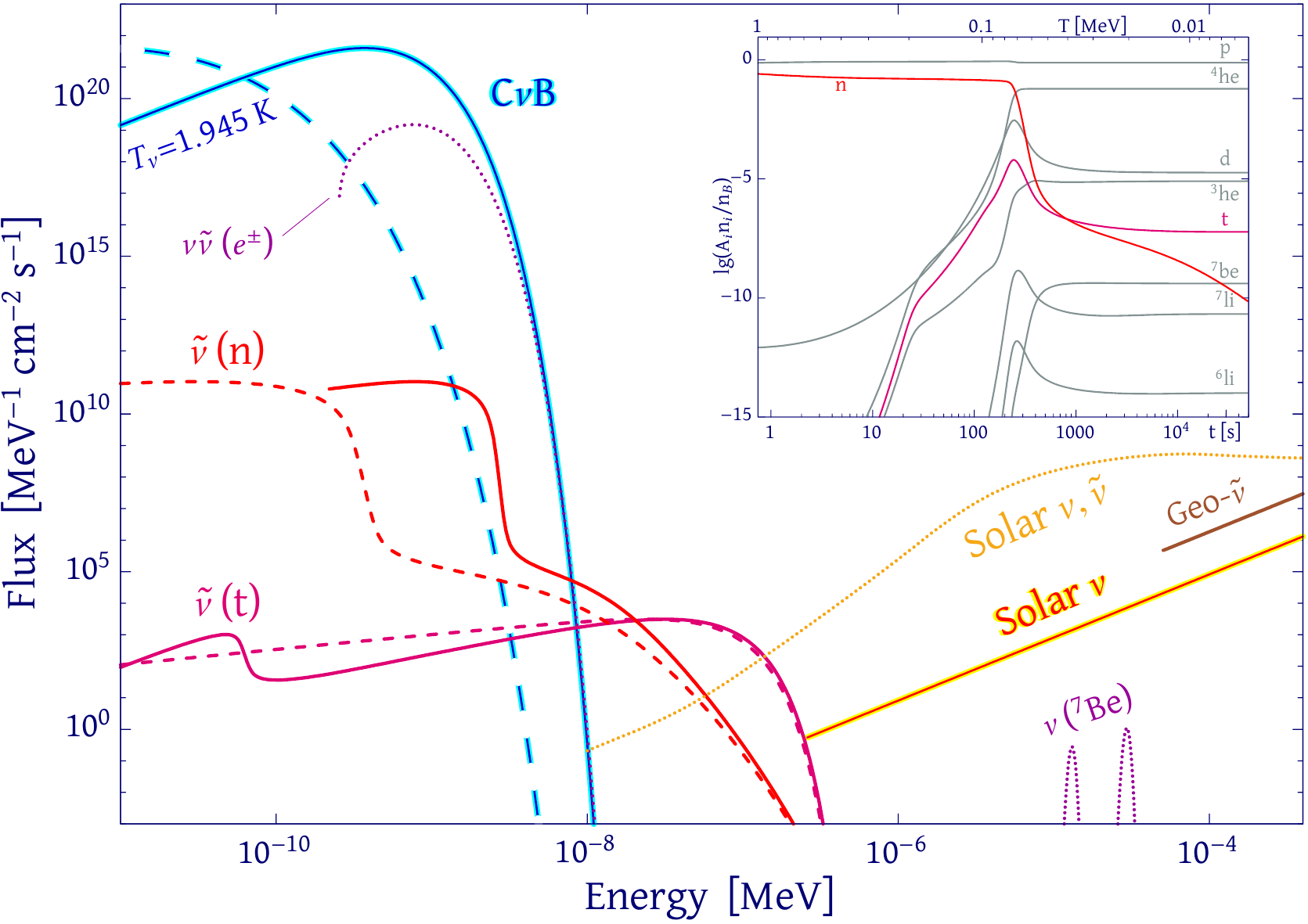}
\caption{\label{fig:antineutrino_spectra} The calculated fluxes of non-equilibrium antineutrinos due to decays of neutrons (red curvatures) and tritons (crimson curvatures) during primordial nucleosynthesis \cite{IY2018} and fluxes of neutrinos and antineutrinos from various natural sources: relic neutrinos and antineutrinos (C$\nu$B) with temperature $T_{\nu}=1.945$ K (see e.g. \cite{W2008}); a non-thermal addition to relic neutrinos and antineutrinos due to $e^+e^-$-annihilation at the beginning of primordial nucleosynthesis \cite{DF1992,DT1992,HM1995,DHS1997}; solar neutrinos from reactions of thermonuclear synthesis \cite{BSB2005}; the calculated spectrum of solar neutrinos and antineutrinos at keV energies emerging from various thermal processes in the solar plasma \cite{VRR}; the doublet of the narrow cosmological neutrino lines from $^7$Be decay \cite{KS2011}; antineutrinos produced geologically (geoneutrinos) in decays of radioactive isotopes \cite{E2006, E2007}. The fluxes in the case of massless neutrinos are represented by dashed lines and in the case of $m_{\nu}=0.01$ eV with solid lines.}
\end{figure}

The total number density of non-equilibrium antineutrinos of energy in the interval between $\tilde{\varepsilon}$ and $\tilde{\varepsilon}+d\tilde{\varepsilon}$ is calculated as the integral of quantity \ref{eq:number_of_particles} diluted due to the expansion of the Universe (where $p$ is expressed through the kinetic energy at the present epoch) over all redshifts:
\begin{equation}
\label{eq:particle_number_density}
\dd n^0_{\nu}(\tilde{\varepsilon})\ =\!\!\!\int\limits_0^{z'(\tilde{\varepsilon})}\frac{\dd n_{\nu}(\tilde{\varepsilon}, z)}{(1+z)^3}\dd z=-\frac{1}{\tau}\!\!\!\int\limits_{0}^{z'(\tilde{\varepsilon})}\!\!\!\frac{n(z)}{(1+z)^3}f^0(\ee)\frac{\dd t}{\dd z}\,\dd z\dd\ee.
\end{equation}
The lower limit on the integral corresponds to the redshift of the present time ($z=0$) and the upper limit $z'(\tilde{\varepsilon})$ equals to ($1/\sqrt{\ee(\ee+2\tilde{m}_{\nu})}-1$) for every fixed value of $\tilde{\varepsilon}$, what is derived from the condition $p^0c{\cdot}(1+z)\leq Q$.
The factor $(1+z)^{-3}$ arises in \ref{eq:particle_number_density} because of dilution of the number density $\dd n_{\nu}(\tilde{\varepsilon}, z)$ due to the expansion of the Universe.
Note, that $f^0(\ee)\propto(1+z)$ (see Appendix A).
The function $\dd t/\dd z$ can be obtained from the relation between time $t$ and temperature $T_{\gamma}(z)=T^0_{\gamma}{\cdot}(1+z)$ during the radiation-dominated era (see e.g. \cite{W2008}):
\begin{equation}
t=1.78\;{\rm s}\cdot\left(\frac{T^0_{\gamma}(1+z)}{10^{10}\,{\rm K}}\right)^{\!\!-2}\!\!+{\rm Const},
\end{equation}
from which it is seen that $\dd t/\dd z<0$, that explains the minus sign in the front of the integral \ref{eq:particle_number_density}.
Thus, the energy spectrum of non-equilibrium antineutrinos at present epoch is
\begin{equation}
\label{eq:particle_distribution_density}
n^0_{\nu}(\varepsilon)=-\frac{1}{Q\tau}\!\!\!\int\limits_{0}^{z'(\tilde{\varepsilon})}\!\!\!\frac{n(z)}{(1+z)^3}f^0(\tilde{\varepsilon})\frac{\dd t}{\dd z}\,\dd z.
\end{equation}
As usual, what is observable is not the number density of particles, but the flux.
The flux of antineutrinos $F(\varepsilon)$ can be calculated by integrating \ref{eq:particle_distribution_density} miltiplied by antineutrino velocity $v(\varepsilon)$ over hemisphere $\Omega/2$:
\begin{equation}
F(\varepsilon)=\frac{v(\varepsilon)}{4\pi}\!\int\limits_{\Omega/2}\!\!n^0_{\nu}(\varepsilon)\cos\Theta\dd\Omega=-\frac{v(\varepsilon)}{4Q\tau}\!\!\int\limits_{0}^{z'(\tilde{\varepsilon})}\!\!\frac{n(z)}{(1+z)^3}f^0(\tilde{\varepsilon})\frac{\dd t}{\dd z}\,\dd z,
\end{equation}
and antineutrino velocity can be found from the relations $v=d\mathcal{E}/dp$ and $\mathcal{E}=\sqrt{(pc)^2+(m_{\nu}c^2)^2}$:
\begin{equation}
\frac{d\mathcal{E}}{dp}=\frac{pc}{\mathcal{E}}\ c=\frac{\tilde{p}}{\tilde{\varepsilon}+\tilde{m}_{\nu}}\ c.
\end{equation}
The calculated fluxes of non-equilibrium antineutrinos are shown in figure \ref{fig:antineutrino_spectra}.
Shapes of these fluxes reflect the temporal evolution of the neutron and triton abundances during primordial nucleosynthesis. Some non-monotonic feature in the low-momentum tail of tritium neutrino spectrum is seen in the case of massless neutrinos and it is absent in the case of massive neutrinos. The reason for this is that for neutrinos with mass equal to $0.01$ eV (which we used in our calculations) this peculiarity would be at energy lower, than $10^{-11}$ MeV, i.e. beyond the range of energy scale of figure 3. Also the high-momentum tail of neutrino spectrum from neutron decays follows the abundance of neutrons during primordial nucleosynthesis, which does not decrease exponentially because of the presence of residual reactions with the participation of neutrons, in which they continue to be produced.

In the calculations we used the following values of lifetimes and released energies: triton's lifetime $\tau_{t}\simeq17.656$ years \cite{AM2005}, neutron's lifetime is $\tau_{n}\simeq$ 880.2 seconds \cite{PDG, S2018}, triton decay energy released is $Q_{t}\simeq18.592$ keV \cite{W2017}; neutron decay energy released is $Q_{n}\simeq782.346$ keV \cite{W2017}.\footnote{Authors in \cite{AM2005} give triton's half life $\tau_{1{/}2}$, which is related to its lifetime as $\tau\!\ln\!2$.}

\subsection{Coulomb interaction}

\begin{figure}[tbp]
\centering 
\includegraphics[width = 1.\textwidth,trim = 0 0 0 0, clip]{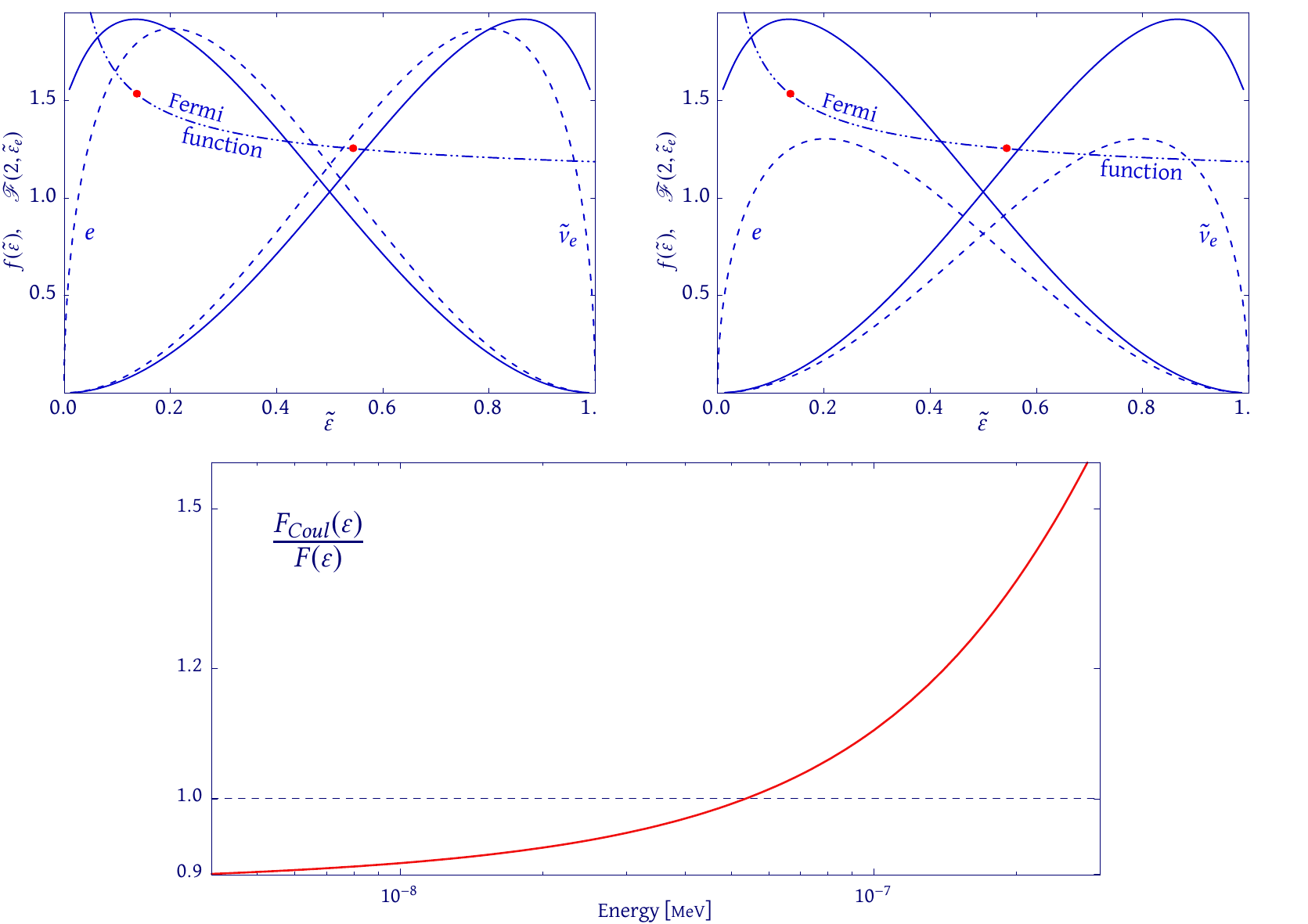}
\caption{\label{fig:deformated_antineutrino_spectra} The upper panels: the energy spectra of an electron and antineutrino emitted in $\beta^-$ decay of triton. Solid lines represent the spectra calculated taking into account Coulomb interaction and dashed lines represent the spectra calculated ignoring Coulomb interaction. Dash-dotted lines are the non-relativistic approximation \ref{eq:fermi_function_approximation} for the Fermi function $\mathscr{F}(Z=2,\varepsilon_e)$ for triton's $\beta^-$ decay. Red circles are the values of the Fermi function calculated relativistically in \cite{BJ1969}. In left figure all the spectra are normalized to unity in accordance with \ref{eq:momentum_distribution_normalizing}, and on right one the spectra differ from each other by the factor of the Fermi function. Note that left panel demonstrates that taking Coulomb interaction into account leads to an increase in the fraction of antineutrinos with higher energies, and the right panel demonstrates that Coulomb interaction effectively leads a lifetime of decaying nuclei to decrease. The bottom panel: the ratio of antineutrino fluxes from decays of tritons calculated taking into account and ignoring Coulomb interaction.}
\end{figure}

Because of the Coulomb interaction between emitted in $\beta^-$ decay electron and the daughter nucleus, the asymptotic momentum of the electron is lower than its momentum at the time of its creation.
By this reason the expression \ref{eq:Fermi's_golden_rule} gives an underestimated value of the transition probability at every momentum.
To correct this discrepancy the Fermi function $\mathscr{F}(Z,\varepsilon_e)$ is used. It is defined as the ratio of squared moduli of the electron wave functions at the site of the nucleus calculated taking Coulomb interaction into account and ignoring it  (see e.g. \cite{K.-K.S.1995, W2004}):
\begin{equation}
\mathscr{F}(Z,\varepsilon_e)=\frac{|\psi^e_{Coul}(0)|^2}{|\psi^e(0)|^2}.
\end{equation}
In calculations of the transition probability the right hand side of equality \ref{eq:Fermi's_golden_rule} is multiplied by the Fermi function.
The Fermi function actually should be calculated relativistically, but there is the following non-relativistic approximation for the Fermi function which works very good in the case of triton decay (because of relatively small electric charge of triton):
\begin{equation}
\label{eq:fermi_function_approximation}
\mathscr{F}(Z,\varepsilon_e)=\frac{x}{1-e^{-x}},\qquad x=\frac{2\pi Z\alpha}{\beta},
\end{equation}
where $Z$ is a charge of the daughter nucleus, $\alpha$ is the fine structure constant, $\beta$ is the velocity of electron in units of the speed of light.
The non-relativistic approximation for the Fermi function $\mathscr{F}(Z{=}2,\varepsilon_e)$ for decay of triton is shown in figure \ref{fig:deformated_antineutrino_spectra}. In addition, values of the Fermi function $\mathscr{F}(Z{=}2,\varepsilon_e)$ which were calculated relativistically in \cite{BJ1969} are put in this figure by red circles to demonstrate the good agreement between the non-relativistic approximation for the Fermi function and the relativistic numerical calculations of the Fermi function.

It was found that the Coulomb interaction increases the fraction of non-equilibrium ``tritium'' antineutrinos of higher energies (see the bottom panel of figure \ref{fig:deformated_antineutrino_spectra}).
In contrast the Coulomb interaction rather slightly modifies the distribution function of antineutrino in $\beta^-$ decay of neutron, it is explained by relatively high energy released ($Q_{n}=782.346$ keV in comparison with $Q_{t}=18.592$ keV).

\subsection{Non-equilibrium antineutrinos as one more independent ``baryometer''}

\begin{figure}[tbp]
	\centering 
	\includegraphics[width=1.\textwidth,trim=0 0 0 0, clip]{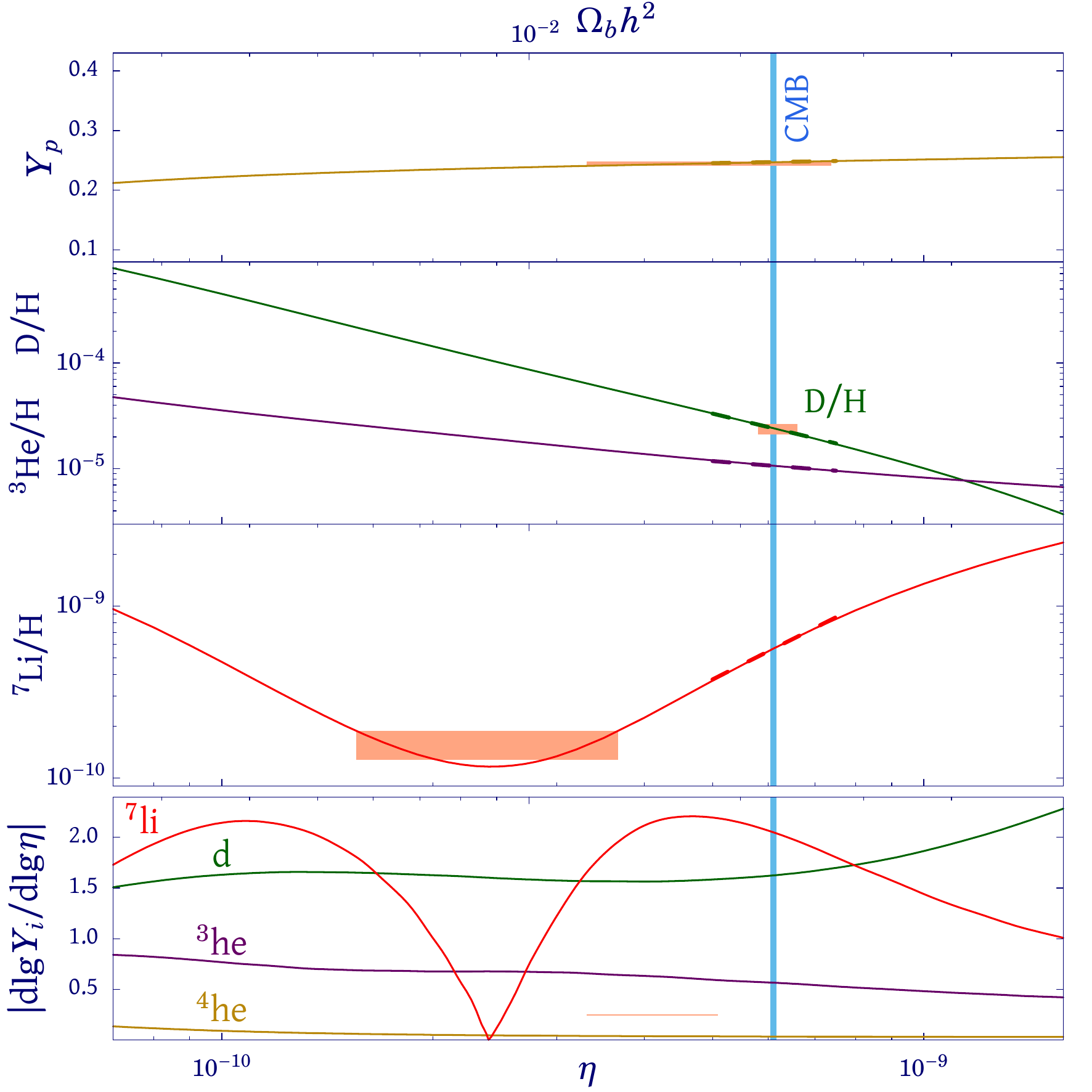}
	\caption{Abundances of light nuclei at the end of BBN depending on $\eta$. Relative sensitivity of light nuclei abundances to change of $\eta$. Vertical line indicates the value of $\eta=\eta^{CMB}$ provided by CMB analysis. Filled areas are error boxes from astronomical observations of light element abundances. Dashed lines are tangent to curvatures at $\eta^{CMB}$.}
	\label{fig:Baryometers}
\end{figure}

\begin{figure}[tbp]
	\centering 
	\includegraphics[width=.475\textwidth,trim=0 0 0 0, clip]{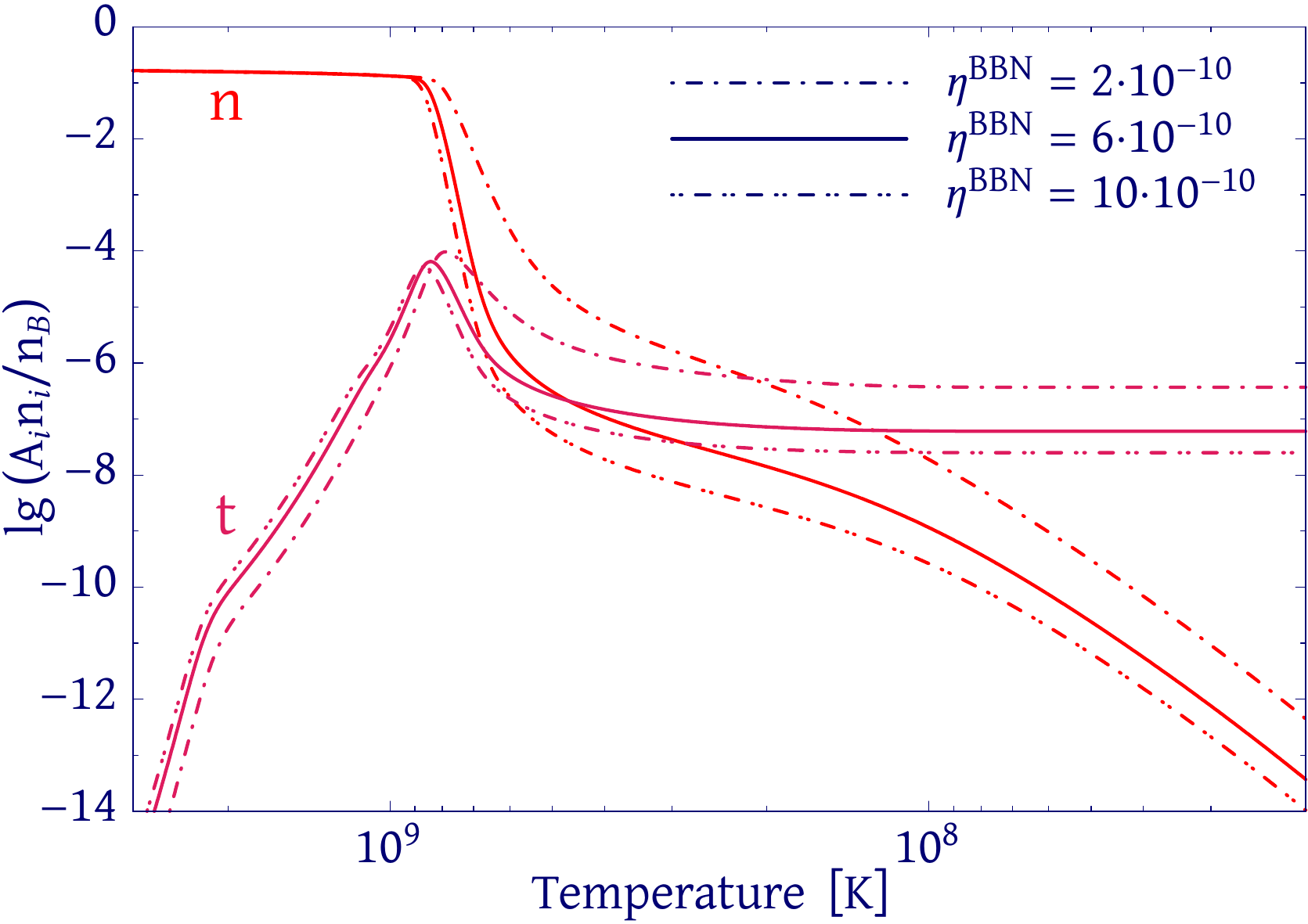}
	\includegraphics[width=.475\textwidth,origin=c,angle=0]{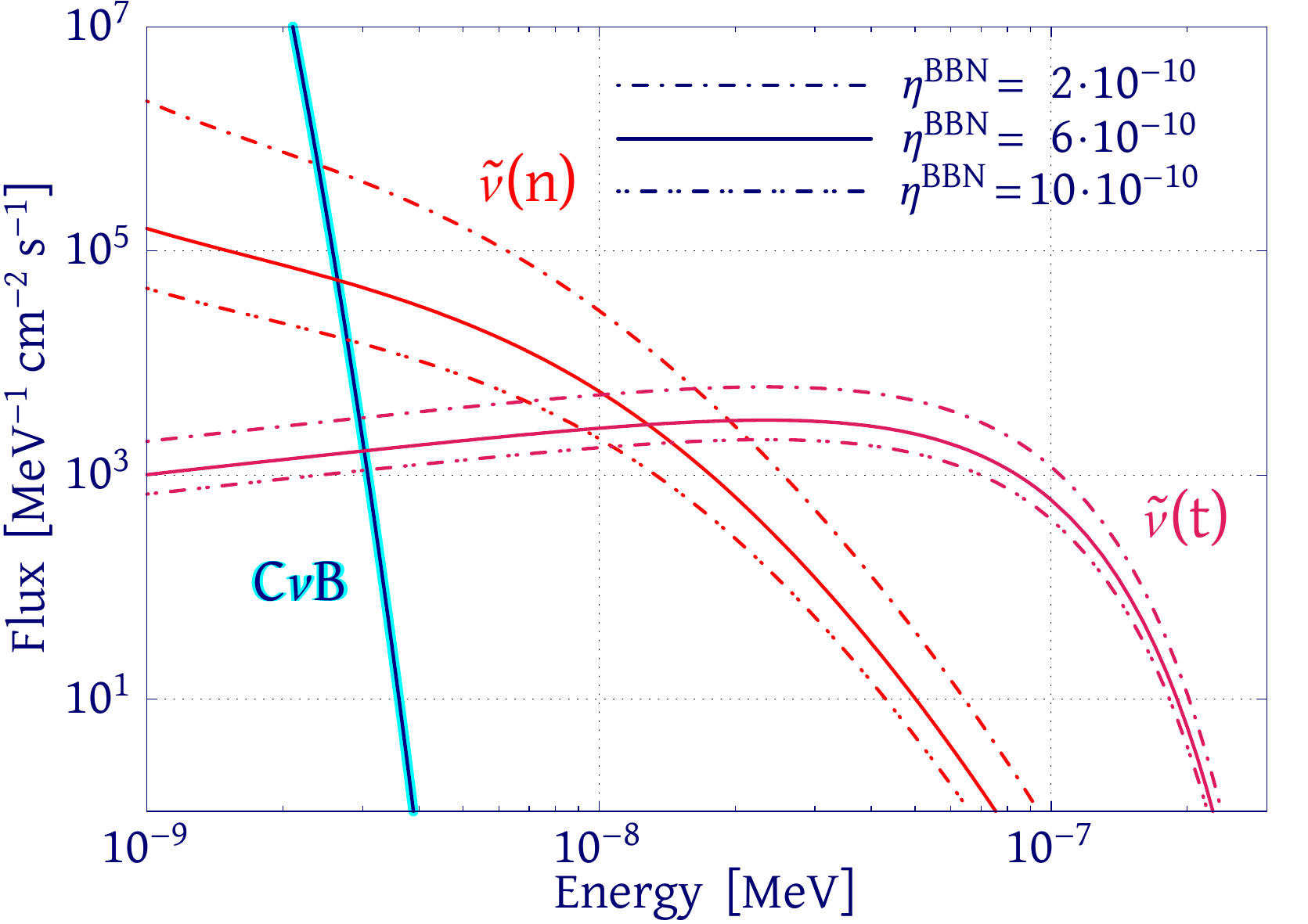}
	\caption{Left panel: the evolution of the mass fractions $A_in_i/n_b$ of neutrons and tritons as a function of the temperature under various values of $\eta^{BBN}$. Right panel: the fluxes of non-equilibrium antineutrinos calculated using the mass fractions of neutrons and tritons which correspond to various values of $\eta^{BBN}$.}
	\label{fig:bario_ratio}
\end{figure}

The only free parameter of primordial nucleosynthesis is the baryon-to-photon ratio $\eta$ (it is convenient to introduce the notation $\eta_{10}\equiv10^{10}{\cdot}\eta$).
Its value is deduced from the comparison of light element observations and the results of BBN numerical calculations.
At present, a good accuracy has been reached in D/H measurements, whose analysis gives in certain cases about one percent determination of the primordial deuterium, D/H=$(2.527\pm0.030){\cdot}10^{-5}$ \cite{CPS2018}.

At the same time a noticeable scatter remains among the results of all these measurements \cite{BZ2016,R-SK2017}.
All available D/H measurements give the unweighted average deuterium abundance of D/H = $(2.53\pm0.16){\cdot}10^{-5}$ \cite{R-SJ2017}.
This estimation corresponds to the range $5.8<\eta_{10}<6.6$ on the assumption of the Standard Model.
Nevertheless, the primordial deuterium is commonly referred to as a baryometer because of its strong dependence on the value of $\eta$.
Meanwhile, $^7$Li has even the more strong dependence on $\eta$ in the range $3.2<\eta_{10}<7.9$ (see figure \ref{fig:Baryometers}), but results of observations and calculations of $^7$Li are differ by a factor of amount $\simeq\,$3 at $\eta^{CMB}$, it is the main statement of so called the lithium problem (see e.g. \cite{F2011}).
Therefore, comparison of current observations and calculations of the primordial light element abundances do not give complete uncontroversial estimates of $\eta$.

Another independent way to determine $\eta$ comes from the analysis of the CMB anisotropies and it gives the value of the baryon density $\Omega_b h^2=0.0224\pm0.0001$ \cite{PC2018} which is related to $\eta^{CMB}$ by $273.77{\cdot}\Omega_bh^2\simeq\eta^{CMB}_{10}$ \cite{CUV2014} and it gives an order of magnitude tighter range of values of $\eta^{CMB}$, in comparison with that shown above: $6.105<\eta^{CMB}_{10}<6.160$.
Note, that this analysis deals with the era of primordial recombination ($\approx380\,000$ years after the Big Bang) and strictly speaking, $\eta^{CMB}$ is not necessarily equal to $\eta^{BBN}$.
Primordial nucleosynthesis and recombination are different cosmological epochs and there is the possibility of a change in $\eta$ on cosmological time scales due to some non-standard processes referred to as physics beyond the Standard Model, such as the decays (or annihilation) of dark matter particles (e.g. \cite{ZI2015} and references therein).
There is the possibility that these processes could also be related to the solution of the lithium problem.
It is worth noting that the overlapping of ranges of $\eta$ coming from analysis of observations of D/H and CMB may point out the approximate constancy of $\eta$ between these epochs.
In this case there is the lithium problem.
But if there were no independent value $\eta^{CMB}$, it would be necessary to refer to ``the deuterium-lithium problem'' meaning the disagreement between ranges of $\eta$ suggested by analyses of D/H and $^7$Li observations.
Moreover, some scenarios have been reported in the literature, they allow one to have quite different $\eta^{BBN}$ and $\eta^{CMB}$ keeping the same abundance of D/H as it is observed (e.g. \cite{YKKMC2014}).

Hence, independent measurements of $\eta^{BBN}$ and $\eta^{CMB}$ are being powerful tool to study physics beyond the Standard Model.
The dependence of non-equilibrium antineutrino spectra on $\eta^{BBN}$ was investigated in this regard.
Right panel of figure \ref{fig:bario_ratio} demonstrates considerable sensitivity of non-equilibrium antineutrino spectra to value of $\eta^{BBN}$.
It suggests that non-thermal antineutrinos potentially could serve as an additional independent baryometer (though it is very challenging problem for present experimental equipment).
The number density of the decaying nuclei depends on $\eta^{BBN}$ both explicitly through the factor $\eta$ and implicitly through the function $X_i(z)$, with the latter sentence being demonstrated in the left panel of figure \ref{fig:bario_ratio}.
The calculations of antineutrino spectra at various values of $\eta^{BBN}=(2,\,6,\,10){\times}10^{-10}$ show that smaller fluxes correspond to higher values of $\eta$ (see the right panel of figure \ref{fig:bario_ratio}).
It is explained by the fact that the more dense matter (higher values of $\eta$) the bigger part of neutrons and tritons tends to transform into $^4$He through nuclear reactions (see the left panel of figure \ref{fig:bario_ratio}).

\section{Conclusions}

The non-equilibrium spectra of antineutrinos of primordial nucleosynthesis have been calculated taking into account the Coulomb interaction between electron and daughter nucleus in $\beta^-$ decay.
It was found that the Coulomb interaction increases prominently the number of ``tritium'' antineutrinos of higher energies.
Namely at the energies $\varepsilon\!>\!10^{-7}$ MeV the ratio of antineutrino fluxes calculated, taking into account and ignoring the Coulomb interaction, amounts to several tens of percents.

Independent measurements of $\eta^{BBN}$ and $\eta^{CMB}$ serves as a powerful tool for studying the physics beyond the Standard Model.
Non-equilibrium antineutrinos may potentially serve as an independent baryometer, which in addition to nuclear baryometers (d and $^7$li) might indicate the value of $\eta^{BBN}$.
In this context, the dependence of non-equilibrium antineutrino spectra on value of $\eta^{BBN}$ is studied.
Calculation of antineutrino fluxes at various values of $\eta^{BBN}$ showed that smaller fluxes correspond to higher values of $\eta^{BBN}$.

The direct detection of the non-equilibium antineutrino background is an observational challenge.
However various possibilities of the direct detection of relic neutrinos are discussed in the literature.
The promising proposals among them are relic neutrino capture on radioactive $\beta$ decaying nuclei \cite{CMM2007} and relic antineutrinos capture on radioactive nuclei decaying via electron capture \cite{CMM2009, LV2011}.
There is also the PTOLEMY project which aims to develop a design for a C$\nu$B detector \cite{PTOLEMY2019}.
For descriptions of other ways and current perspectives of the direct detection of the relic neutrinos/antineutrinos see, for instance, the following detailed works \cite{R2009, YFL2015, BR2015} as well as \cite{XZ2011} and references therein.
Note, that the difference between capture rates of non-equilibrium antineutrinos in cases of Majorana and Dirac neutrinos depends on the value of the lightest neutrino mass and on the type of mass hierarchy \cite{Roulet_2018}. The difference in detection rates of Majorana and Dirac neutrinos became significant only for non-relativistic neutrinos. Hense, in the range [$10^{-8}$, $10^{-7}$] MeV, where non-equilibrium antineutrinos dominate it will not be reveal itself for antineutrinos with masses less than 0.01 eV, because such antineutrinos are still relativistic (see Fig. 3). Non-equilibrium antineutrinos with momentums less than $10^{-8}$ MeV where they become non-relativistic are dominated by C$\nu$B neutrinos (the C$\nu$B flux larger by several orders of magnitude).

Distortions in the spectrum of relic neutrinos can provide very useful observational information about cosmological epochs which immediately follow neutrino decoupling.
This information could be used to extract values of a number of cosmological parameters and set constraints on various cosmological scenarios.\\

\noindent
\textit{Acknowledgments}
We thank Edoardo Vitagliano for helpful comments and we also thank the anonymous referee for careful reading of our manuscript and valuable comments and suggestions.
The work has been supported by Russian Science Foundation (Grant No. 18-12-00301).

\appendix

\section{The antineutrino momentum spectrum in $\beta^-$ decay}
The antineutrino momentum spectrum in $\beta^-$ decay is calculated using Fermi's golden rule (see  e.g.  \cite{K.-K.S.1995, W2004}):
\begin{equation}
\label{eq:Fermi's_golden_rule}
\dd\mathcal{W}=w(p_{\nu})\dd p_{\nu}=\frac{2\pi}{\hbar}|\langle n|H_{\beta}|m\rangle|^2\dd\rho(Q),\qquad \dd\rho(Q)\equiv \left.\frac{\dd n_e\dd n_{\nu}}{\dd\hspace{-.1em}\mathcal{Q}}\right|_{\mathcal{Q}=Q}.
\end{equation}
Here $\dd\mathcal{W}$ is the probability per unit time of the $\beta^-$ transition from the state in which there is a $\beta^-$ unstable system to a state in which there are an antineutrino of momentum in the interval between $p_{\nu}$ and $p_{\nu}+dp_{\nu}$ and an electron of momentum in the interval between $p_e$ and $p_e+\dd p_e$; $\langle n|H_{\beta}|m\rangle$ is the matrix element of the Hamiltonian operator
of the weak interaction between the initial state |m\textrangle\ and the final state |n\textrangle,  $\dd\rho(Q)$ is the number of the possible final states (for the given antineutrino momentum) per unit energy range around the energy released $Q$ in $\beta^-$ decay, where the product of infinitesimal phase spaces of the electron and the antineutrino is
\begin{equation}
\dd n_e\dd n_{\nu}=\frac{V\dd^3p_e}{(2\pi\hbar)^3}\frac{V\dd^3p_{\nu}}{(2\pi\hbar)^3}.
\end{equation}

As (i) the matrix element $\langle n|H_{\beta}|m\rangle$ weakly depends on energy, (ii) the recoil energy of the daughter nucleus is negligible compared with the energy taken by the electron and the antineutrino in the form of their kinetic energy (that gives $\varepsilon_e+\varepsilon_{\nu}=\mathcal{Q}$), and (iii) $p_e=\frac{1}{c}\sqrt{\varepsilon_e(\varepsilon_e+2m_ec^2)}$ for the emitted electron and $p_{\nu}=\varepsilon_{\nu}/c$ for the emitted antineutrino, then collecting all constant factors into the only constant $\mathcal{C}'$ we obtain the shape of the probability of finding the antineutrino with the momentum $p$:
\begin{equation}
w(p)=\mathcal{C'\!F}(p),
\end{equation}
with
\begin{equation}
\label{eq:shape}
\mathcal{F}(p)=\frac{1}{c^3}\sqrt{(Q-pc)(Q-pc+2m_ec^2)}(Q-pc+m_ec^2)p^2.
\end{equation}
This expression is analogous to well known result for the transition probability for an electron emitted in $\beta^-$ decay with momentum $p_e$:
\begin{equation}
\label{eq:shape_e}
\mathcal{F}(p_e)=\frac{1}{c^3}(Q-\sqrt{p_e^2c^2+m_e^2c^4})^2p_e^2.
\end{equation}
Dependencies \ref{eq:shape} and \ref{eq:shape_e} are shown in figure \ref{fig:deformated_antineutrino_spectra}.

The transition probability \ref{eq:Fermi's_golden_rule} integrated over all possible values of the antineutrino momentum gives the total probability of $\beta^-$ decay per unit time and it is equal to the decay rate $\lambda$ or the inverse mean lifetime $\tau$ of decaying nucleus:
\begin{equation}
\label{eq:whole_probability}
\mathcal{W}=\int\limits_0^{Q/c}w(p)dp=\int\limits_0^{Q/c}\mathcal{C'\!F}(p)dp=\lambda=\frac{1}{\tau}.
\end{equation}
It is seen from the normalization conditions \ref{eq:momentum_distribution_normalizing} and \ref{eq:whole_probability}, that the relation between the momentum spectrum $f(p)$ of antineutrino emitted in $\beta^-$ decay \ref{eq:number_of_particles} and the function $\mathcal{F}(p)$ is the following:
\begin{equation}
\label{eq:distribution_function}
f(p)=\tau\mathcal{C'\!F}(p).
\end{equation}
It is convenient to take \ref{eq:distribution_function} in the dimensionless form using dimensionless parameters: $\pp=pc/Q$, which takes values from 0 to 1, $\tilde{m}_e=m_ec^2/Q$, $\mathcal{C}=\tau\mathcal{C}'Q^4/c^5$, that gives
\begin{equation}
\label{eq:distribution_function_dimensionless}
f(\tilde{p})=\mathcal{C}\sqrt{(1-\pp)(1-\pp+2\tilde{m}_e)}(1-\pp+\tilde{m}_e)\pp^2.
\end{equation}

\section{The evolution of the antineutrino spectrum in the expanding Universe}
The momentum of particles propagating freely through the Universe obeys the following relation (see e.g. \cite{GR2018, KT1990}):
\begin{equation}
\label{eq:momentum_low}
p(t)a(t)=p^ia^i,
\end{equation}
where $a(t)$ is the scale factor of the Universe related to the cosmological redshift $z(t)$ through the definition $1+z(t)=1/a(t)$ (and $a=1$ at the present epoch).
The superscript $i$ stands for ``initial'' values.

It is seen from the relation \ref{eq:momentum_low} that in the expanding Universe the momentum of antineutrinos decreases as time passes.
Due to the above fact antineutrinos become non-relativistic at some moment, and the relation between its momentum and kinetic energy $\varepsilon$ takes the form
\begin{equation}
p^0c=\sqrt{\varepsilon(\varepsilon+2m_{\nu}c^2)},
\end{equation}
where $m_{\nu}$ is the antineutrino mass.
The superscript 0 stands for values taken at the present epoch ($z=0$, $t_0=13.8$ Gyr).
It can be put in the following dimensionless form:
\begin{equation}
\pp^0=\sqrt{\ee(\ee+2\tilde{m}_{\nu})},
\end{equation}
where $\ee=\varepsilon/Q$ and $\tilde{m}_\nu=m_{\nu}c^2/Q$.

Substituting the relation \ref{eq:momentum_low}, taken in the form $p^i=p^0/a^i=p^0\cdot(1+z^i)$, into the expression \ref{eq:distribution_function_dimensionless} one obtains the fraction of antineutrinos of momentum in a given interval, which were produced at fixed value of $z=z^i$:
\begin{equation}
f(\pp^i)\dd\pp^i=f(\pp^0{\cdot}(1+z^i))(1+z^i)\dd\pp^0=f^0(\pp^0)\dd\pp^0.
\end{equation}
It corresponds to the fraction of antineutrinos, which at present have kinetic energy in the interval from $Q\cdot\ee$ to $Q\cdot(\ee+\dd\ee)$:
\begin{equation}
f^0(\ee)\dd\ee=f^0(\pp^0(\ee))\frac{\dd \pp^0(\ee)}{\dd\ee}\dd\ee.
\end{equation}

\section*{References}

\bibliography{mybibfile}

\end{document}